\documentclass[12pt,preprint]{aastex}

\shorttitle{Massive Clusters in Arp 220}
\shortauthors{Wilson et al.}

\begin{document}

\title{Two Populations of Young Massive Star Clusters in Arp 220}


\author{Christine D. Wilson\altaffilmark{1}, William E. Harris, Rebecca Longden}

\affil{Department of Physics \& Astronomy, McMaster University,
Hamilton, Ontario, Canada L8S 4M1}

\and

\author{N. Z. Scoville}
\affil{Caltech 105-24, Pasadena CA, U.S.A. 91125}

\altaffiltext{1}{Also Smithsonian Astrophysical Observatory,
Submillimeter Array Site, Hilo, HI 96720}

\begin{abstract}
We present new optical observations of young massive star
clusters in Arp 220, the nearest ultraluminous infrared galaxy,
taken in $UBVI$ with the Hubble Space Telescope ACS/HRC camera.
We find a total of 206 probable clusters whose spatial distribution
is centrally concentrated toward the nucleus of Arp 220.
We use model star cluster tracks to determine
ages, luminosities, and masses 
for 14 clusters with complete $UBVI$ indices or previously 
published near-infrared data.
We estimate rough masses for 24 additional clusters 
with $I < 24$ mag from $BVI$ indices alone.
The clusters with useful ages fall
into two distinct groups: a ``young'' population ($< 10$ Myr) 
and an intermediate-age population ($\simeq 300$ Myr).  
There are many clusters 
with masses clearly above $10^6 M_{\odot}$ and possibly
even above $10^7 M_{\odot}$ in the most extreme instances.  
These masses are high enough that the clusters being formed in the Arp 220
starburst can be considered as genuine young globular clusters.
In addition, this study allows us to extend the observed correlation between
global star formation rate and maximum cluster luminosity by more
than an order of magnitude in star formation rate.

\end{abstract}



\section{Introduction}

Young massive star clusters \citep[or YMCs; see][]{larsen02}
are an intriguing mode of star formation
in the present-day universe. While their older and usually more massive counterparts,
the classic globular clusters, are found around almost every type of galaxy
\citep{h01}, 
rich populations of luminous blue star clusters are found 
predominantly in starburst and interacting systems
\citep[][among many others]{h92,w93,ws95,bastian05}. 
However, the fact that individual young clusters have been
found in several nearby dwarf galaxies 
\citep[e.g.][]{c94,o94,bhe02} 
and small populations are found in several nearby spiral galaxies
\citep{m96,l00,larsen01,larsen02} 
suggests that massive star cluster formation is a relatively 
wide-spread phenomenon, although it seems to occur with
high efficiency only in the most active star-forming systems.

Many questions remain about 
the properties and ultimate fate of these young massive clusters.
The combination of high and variable reddening and uncertain
ages has often made it difficult to determine accurate masses for them. 
Dynamical masses are the most reliable, but these are only available for
a few systems \citep{ho96,mengel02}. Recently, 
intermediate-age clusters with dynamical masses 
greater than 10$^7$ M$_\odot$ have been identified in
two merger-remnant galaxies, \object{NGC 7252}
and \object{NGC 1316} \citep{maraston04,bastian06}.
Among merger and merger-remnant galaxies, 
only in the nearest system, the Antennae 
(\object{NGC 4038}\object[NGC 4039]{/39}), have accurate photometric
masses and ages 
been determined for large numbers of clusters \citep{w99,zf99,wz02}.
In the Antennae, both the youngest clusters ($< 6$ Myr) 
and a slightly older population of clusters
(25-160 Myr) reach masses as large as $3-4\times 10^6$ M$_\odot$
\citep{zf99,wz02}. 
A single YMC in \object{NGC 6946}, a much more modest starburst
system, has a mass near $10^6 M_{\odot}$ as well \citep{larsen01}.
In comparison, the most massive globular clusters
range from $5\times 10^6$ M$_\odot$ for $\omega$ Cen in 
the Milky Way \citep{meylan95}
to above $\sim 10^7$ M$_\odot$ for the most extreme known cases
such as the cluster G1 in \object{M31} \citep{meylan01},
the most massive clusters in \object{NGC 5128} \citep{mar04}, and the most
luminous globular clusters in supergiant elliptical
galaxies \citep{h05}.
Thus, an intriguing question is whether we can identify very young 
clusters ($< 10$ Myr) as massive as $10^7$ M$_\odot$ in 
galaxies in the local universe.
Since stars and star clusters form in molecular gas, the best
place to search for the most massive young star clusters is in
the most gas-rich galaxies.

\object{Arp 220} is the closest example of an ultra-luminous infrared galaxy
\citep{so87}. 
At a distance of 77 Mpc, it is only four times more distant than
the Antennae system and only slightly more
distant than the merger remnant NGC 7252, 
and represents our best chance to identify
and study very young massive star clusters in an ultra-luminous
infrared galaxy. Arp 220 has
faint tidal tails and distortions seen in both optical and
HI emission in the outer parts of the galaxy 
\citep{a66,j85,h00} 
and twin nuclei separated by only 300 pc \citep{s98,s99,so99}.
By comparison with the models of \citet{mh96} and assuming the
two progenitor galaxies to be similar to the Milky Way,
\citet{m01} estimate the time since the beginning of the interaction 
to be $\sim 700$ Myr, with the recent burst of star formation
that powers the galactic superwind and bubbles likely to have started
10-100 Myr ago.
Arp 220 contains a molecular gas mass of
$9\times 10^9$ M$_\odot$ \citep{s97}.  Although this is
roughly half the total mass of molecular gas in the
Antennae \citep{g01}, its molecular gas is concentrated to the inner 750 pc
radius of the galaxy, so that its average molecular gas surface density
reaches an astounding $5\times 10^4$ M$_\odot$ pc$^{-2}$ \citep{s97}.
This surface density corresponds to $A_v = 3300$ mag for a standard
gas-to-dust ratio and is 
comparable to the average surface density in a dense
star-forming core inside a giant molecular
cloud \citep{motte01}. 
In short, if any nearby galaxy has the fuel and the conditions to 
be forming extremely massive young star clusters, it should be Arp 220.

The first observations of Arp220 with the Hubble Space
Telescope identified eight compact objects, of which the two
brightest were suggested to be massive associations
of young stars \citep{s94}. 
Near-infrared observations by \citet{s98} 
identified eight bright star cluster candidates in Arp 220. 
\citet{s01} combined these two sets of data to estimate ages for
three of these star clusters in the range of 10-100 Myr.
In this paper, we present new $UBVI$ observations 
obtained with the Hubble Space Telescope (HST) to search for
additional young massive star clusters in Arp 220 and determine more
detailed properties for them.
The observations and data reduction are presented in \S~\ref{s2}, and we
estimate masses, ages, and reddenings for the cluster candidates
by comparison to the \citet{bc03} models in \S~\ref{s3}.
We discuss the implications of our results for the formation
of young massive clusters in \S~\ref{s4}.

\section{Observations and Data Reduction\label{s2}}

\begin{figure}
\plotone{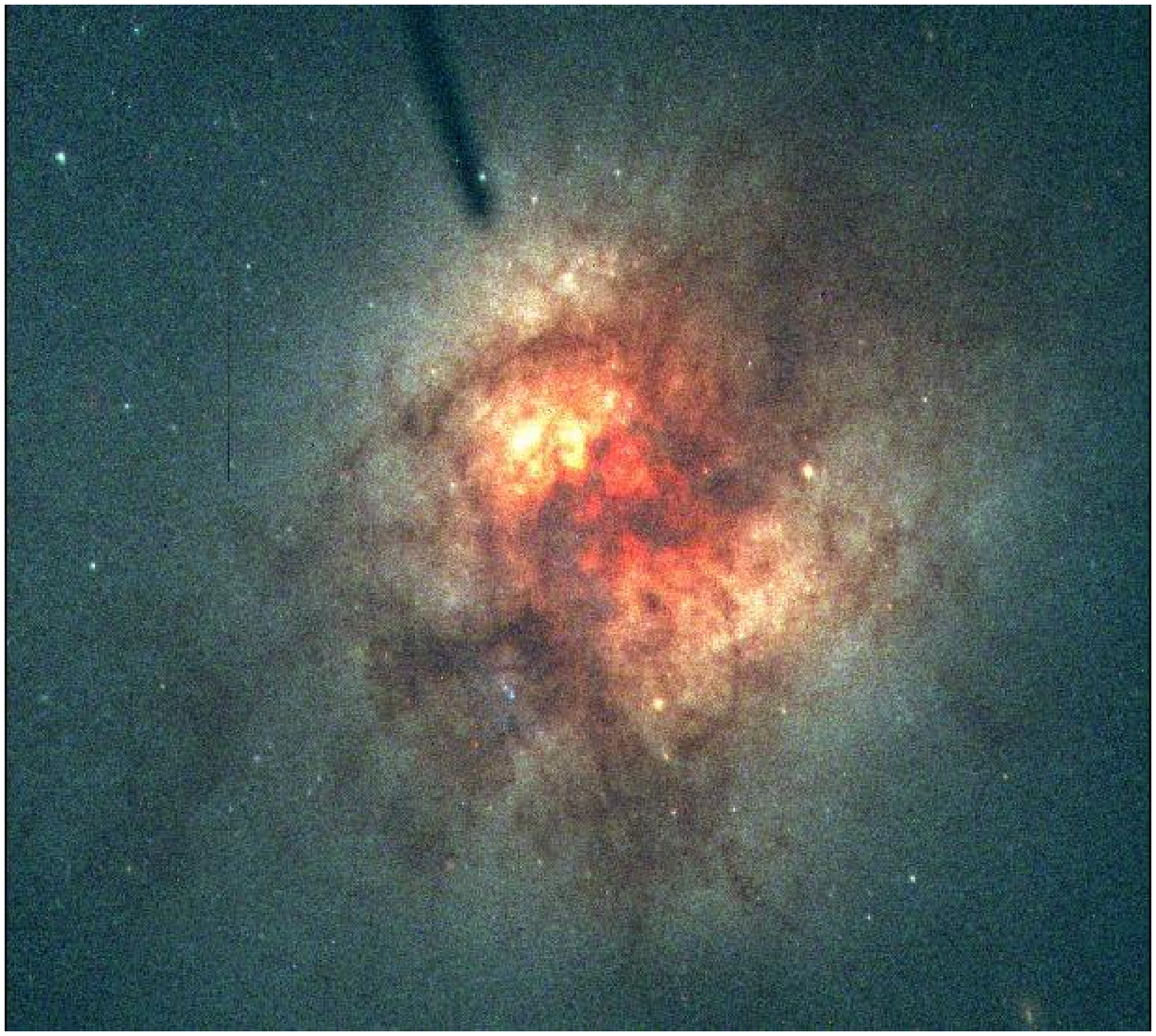}
\caption{Three-color image of Arp 220 produced using the F435W,
F555W, and F814W exposures. This image has not been corrected for
geometric distortion and has a position angle of 77 degrees,
so that north is roughly to the left and east is down.\label{fig-color}}
\end{figure}

Our new observations of Arp 220  were obtained with the Advanced Camera for
Surveys (ACS) on 2002 August 11 through its High Resolution Channel (HRC),
which has a field of view of $26''\times 29''$ and a scale of
$0\farcs027$ per pixel. The total integration time was 5460 s in F330W, 1240 s
in F435W, 1200 s in F555W, and 2640 s in F814W.
We obtained four exposures in each filter to facilitate
the removal of cosmic rays.  The Multidrizzled versions of the images
were downloaded from the STScI Archive, providing
averaged, cosmic-ray-cleaned, and astrometrically rectified images
in each filter to work with.  A color image of our field centered on
Arp 220, as constructed from the three longer-wavelength
filters ($BVI$) is shown in Figure~\ref{fig-color}.

Globular clusters
and YMCs have typical half-light diameters near $\sim 5$ pc, which, 
at the 77 Mpc
distance of Arp 220, correspond to diameters less than $0\farcs02$. These
diameters are 
much smaller than the stellar point spread function (PSF) 
diameter, which on our Multidrizzled frames
is near 3.0 pixels ($0\farcs08$) full-width at half-maximum (FWHM).  
Thus, accurate PSF fitting
can be performed.
The photometry was carried out
with DAOPHOT \citep{s87} in its most recent standalone version
{\sl daophot-4}.
We employed the normal sequence of finding starlike
objects, carrying out small-aperture photometry on them,
defining the stellar point-spread function from selected bright,
uncrowded objects on the image, and finally fitting the PSF to
all starlike objects detected in each filter with the standalone
{\sl allstar} code.  

To find objects for photometry, we constructed a fiducial $(B+V+I)$ image
by summing the Multidrizzled F435W, F555W, and F814W frames.  All starlike or
near-starlike objects clearly visible by eye inspection on this
summed image were then marked for aperture photometry in each filter and 
subsequent PSF fitting through {\sl allstar}.  
(We experimented with various automated object detection procedures,
such as through {\sl daophot/find} with normal threshold levels, but these led to large
numbers of false detections in the many regions where the background light
was strongly variable over small spatial scales. In the end, we had to cull
these lists by eye inspection and thus we ended up using the lists determined
by direct visual examination.)  The limiting magnitude
of our photometry differs quite
significantly from place to place on the images because of the strongly
variable background light. 
However, even though there is no single limiting magnitude that can be
applied uniformly across the whole field,
we expect our photometry to be complete across 
most of the area to $I \simeq 25.0$ mag;
our discussion in \S\ref{s3} and \S\ref{s4}
relies primarily on the brighter objects with $I < 24$ mag.
Our final list contains 206 objects visible
in F814W, not all of which are visible in the other filters.

The PSF fitting radius we used in all four filters was 3 pixels.
The PSF-fitted instrumental magnitude will differ from the
aperture magnitude through the standard $0\farcs5$-radius aperture
required for standardization of the ACS magnitude scales.
Because of the variable background in the images and the faintness
of the sources, we could obtain good empirical aperture corrections
for only the F814W image. To work around this problem, we used the tables in
\citet{s05}
to calculate aperture corrections from a radius of 3 pixels to 
a radius of $0\farcs5$, and then from there to ``infinite'' radius following
their prescription.  Then, we added 
these values to the small measured offset between our
PSF magnitudes and those measured
in an aperture with a radius of 3 pixels.  Finally, we added the
filter zeropoints to the VEGAMAG system given in \citet{s05} to set the final magnitudes.
No further color terms were added to these final filter-based magnitudes, since
the necessary $UBVI$ color indices were not available, or not precisely enough known,
for many objects in the list and thus could not have been calculated in a homogeneous
way.  Adopting some mean color index would also have been invalid since the
actual object-to-object range in color is large here.
However, the color terms are small in our color range of interest,
particularly for $BVI$ \citep{s05}.
As will be seen from the two-color graphs shown later, the point-to-point differences in the reddening and 
background light introduce a large enough degree of scatter in the
deduced intrinsic colors of the objects to make any such residual corrections unimportant
in the later analysis.
The values of the calibration parameters are given in Table~\ref{tbl-cal}.

Initial coordinates for each detected cluster were calculated using the 
astrometric header information in the F814W image files. Since the absolute pointing of
HST can be off by $\sim 1^{\prime\prime}$ or more \citep[e.g.][]{wz02},
we compared our coordinates for seven clusters that we were able to
cross-identify in common with the near-infrared $HST/NICMOS$ study of
\citet{s98}, in which the coordinates were established to well within
$1''$ absolute accuracy from the position of the central radio source
(see their discussion).
We found that it was necessary to shift the raw ACS coordinates
by $3\farcs5$ in right ascension and $0\farcs3$
in declination to bring them onto the
same system as that of \citet{s98} precessed to J2000, which we have adopted.

\begin{figure}
\plotone{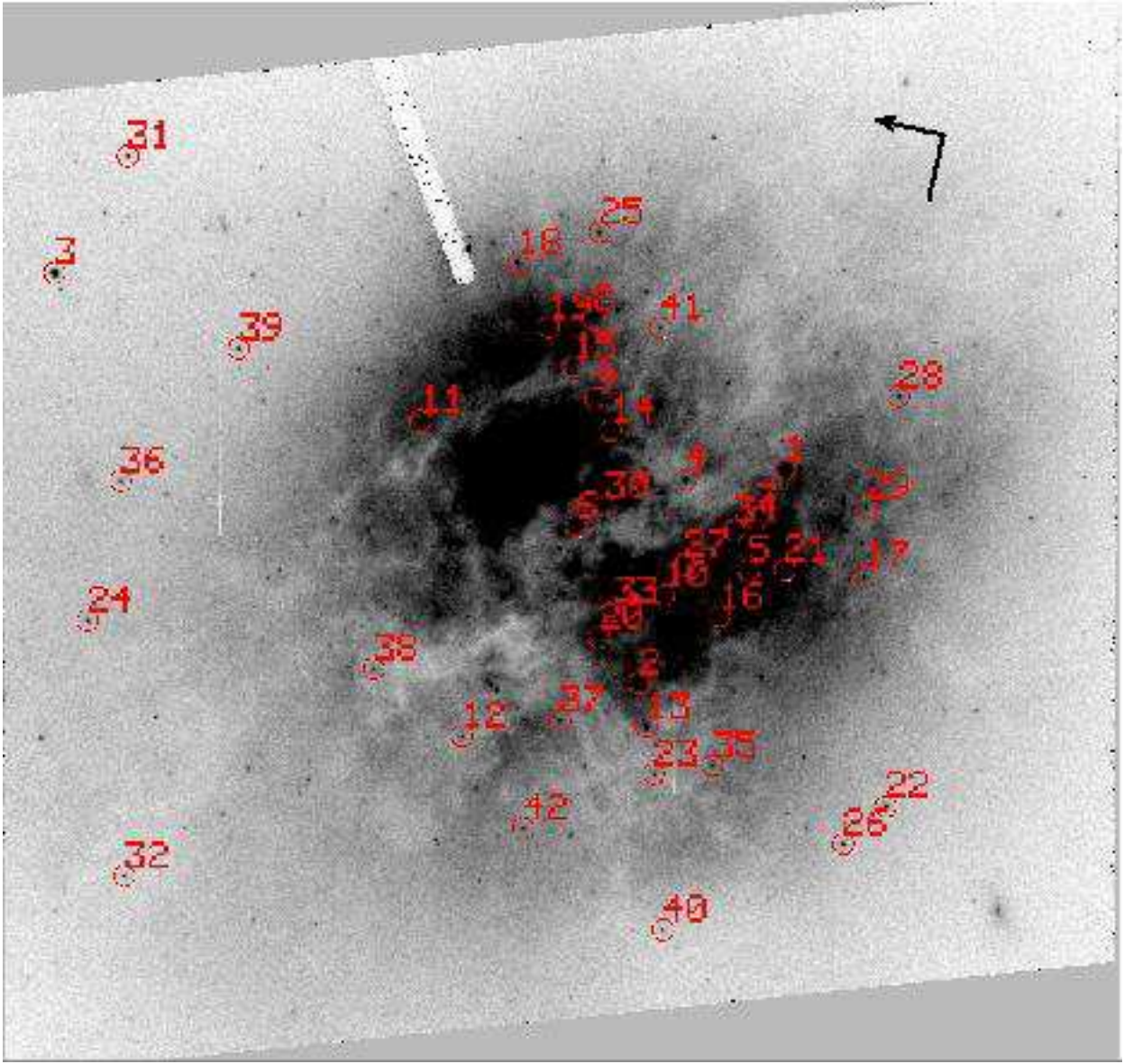}
\caption{Individual cluster candidates brighter than $I=24.0$ mag
in the Arp~220 field are identified on this fiducial image, which is
the sum of the (F435W+F555W+F814W) Multidrizzled frames. There
are 42 marked objects.  The directions North and East are marked
at upper right (North is the arrowtip pointing to upper left,
East points to lower left.) \label{outerfield}}
\end{figure}

\begin{figure}
\plotone{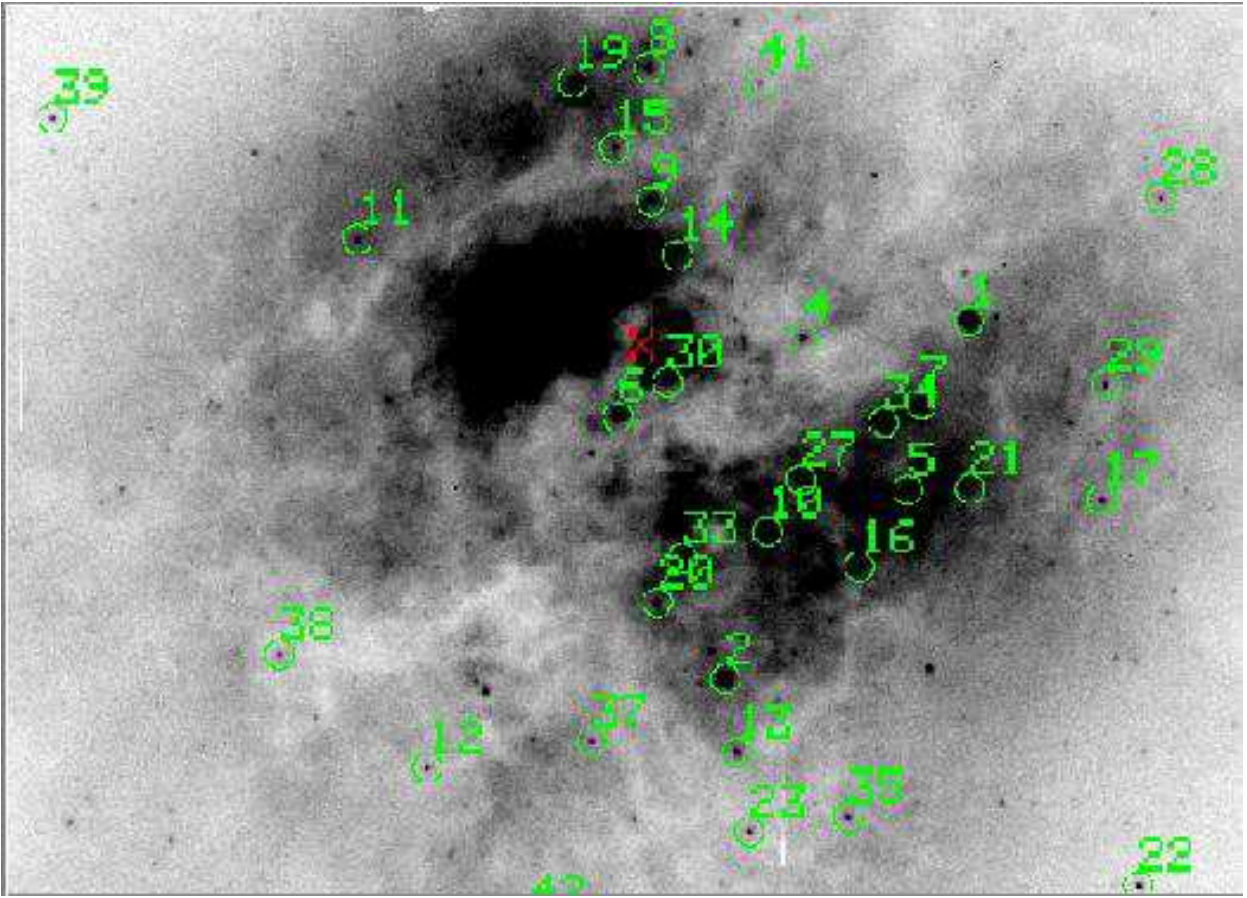}
\caption{Individual cluster candidates brighter than $I=24.0$ mag
in the central part of the Arp~220 field are identified.
Orientation is the same as in the previous figure. 
The cross in red near the center marks the central position
of the galaxy adopted by Scoville et al. (1998). \label{innerfield}}
\end{figure}

\begin{figure}
\includegraphics[angle=-90,scale=.8]{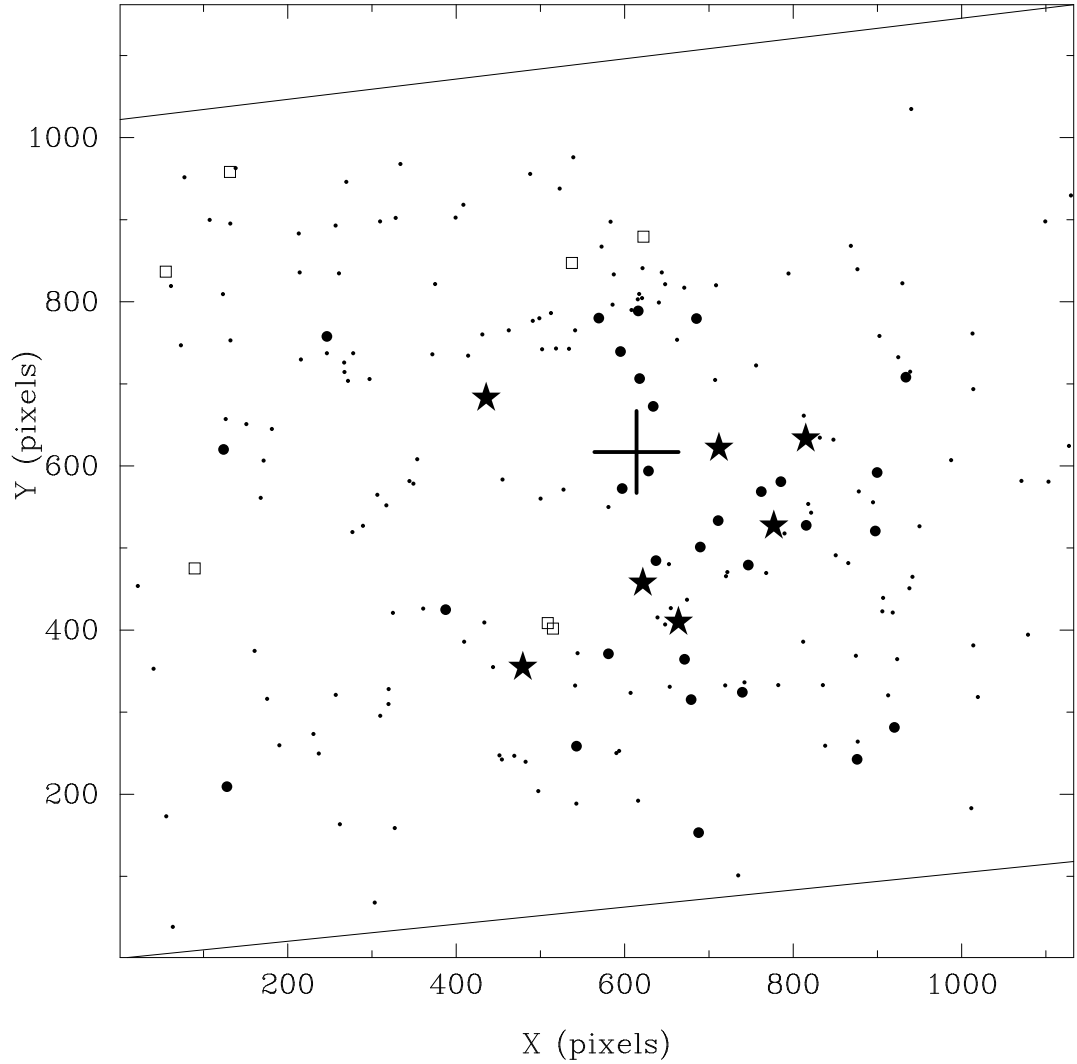}
\caption[f4.eps]{Positions of measured objects in the Arp 200 field.
The large {\sl solid cross} marks the Scoville et al. (1998) 
galaxy center; {\sl solid stars} are objects with near-infrared
photometry from Scoville et al.; {\sl open squares} are objects
with measurements in all four optical bands $UBVI$; 
{\sl large dots} are other objects brighter than $I = 24.0$ mag;
and {\sl small dots} are objects fainter than $I = 24.0$ mag. \label{xyplot}}
\end{figure}

The central region of Arp 220 imaged in our ACS/HRC data is shrouded in
large amounts of dust. Thus, few clusters were visible
through the $U$ filter (F330W) despite our long exposure time.
By comparing the photometry lists from the different filters, we find that
there are only 7 clusters detected in all four filters ($UBVI$) out of
a total of 206 objects detected in F814W ($I$) with brightnesses
$< 26$ mag.  Most of the
remaining clusters were also visible in F435W and F555W ($B$ and $V$).
The final coordinates and $BVI$ magnitudes for all the clusters 
are given in Table~\ref{tbl-phot}, where the measured objects are
numbered in order of brightness in $I$.
For purposes of field identification, we mark the brightest 42 of these
(with $I < 24.0$ mag) in the finder charts of
Figures~\ref{outerfield} and \ref{innerfield}.
The locations of all measured objects, coded by which types of photometric
data are available for them, are shown in Figure \ref{xyplot}.

The expected contamination of our sample due to Milky Way foreground
stars is negligible;  generic starcount models
\citep[e.g.][]{bahcall1984}, as well as direct starcounts from the
Hubble Deep Field and Medium Deep Fields \citep[e.g.][]{san1996,men1996}
predict that we should expect to see less than one foreground star with
$I<26$ mag within
the HRC field size of 0.2 arcmin$^2$. Similarly, the galaxy counts from the
Hubble Deep Field \citep{w96} suggest we should see at most 6 galaxies 
with $I < 26$ mag in our field. Most of these galaxies 
would likely be significantly non-stellar and the large reddening intrinsic
to Arp 220 would further reduce the background galaxy counts in our image.

\section{Cluster Masses and Ages\label{s3}}

For a strong starburst environment like Arp 220, the observed colors of
the embedded star clusters can be strongly affected by large differences in
both cluster age and reddening internal to the galaxy.
Appropriate single-age stellar population models are thus a key to
interpreting the observations.  To help take a first step toward
understanding the cluster age distribution and thus the times of the
major recent starbursts, we
have estimated cluster ages and reddenings by comparing the colors of
individual star clusters to the models of \citet{bc03}, and from these we
derive their dereddened luminosities and hence masses. We adopt
a \citet{s55} initial mass function\footnote{Adopting
the \citet{c03} initial mass function instead of the Salpeter law would 
decrease the
cluster masses derived here by about a factor of two, because of the 
significantly lower numbers of stars on the lower main sequence.}, 
and the \citet{ccm} reddening
law with $R = 3.1$. We also correct for Galactic foreground extinction 
of $E(B-V) = 0.036$ \citep{bh}. 
Given the difficulties noted above in transforming the measured magnitudes
to the standard $UBVI$ system, a slightly
better procedure would be to employ models specifically calculated to give 
colors in the natural HST/ACS filter system.
However, we find that 
the groups of cluster ages in Arp 220 are sufficiently distinct
to allow very useful conclusions with the present analysis (see below).
%For the present, the results
%we find for the cluster ages are distinct enough to allow
%very useful conclusions about their age range (see below).

\citet{wz02} have noted that H$\beta$ contamination
can have a significant effect on the observed $V$ flux for clusters
with ages between 1 and 5 Myr. In their study of
the Antennae, \citet{wz02} used the H$\alpha$ image
to estimate and correct for this contamination.
Unfortunately, there is no high-resolution H$\alpha$ image available
for Arp 220, although ground-based integral-field spectroscopy by
\citet{c04} shows extensive H$\alpha$ emission covering much of
the galaxy center. In the analysis below, we 
have not attempted to correct our $V$ magnitudes
for any line contamination.

The exact procedures adopted for each cluster depend upon the range of
color indices available. Since the uncertainty in the masses and ages
differs quite substantially from one cluster to the next, the
different classes of clusters are discussed in more detail below.

\subsection{Clusters with $U$ photometry\label{opt}}

\begin{figure}
\includegraphics[angle=-90,scale=.7]{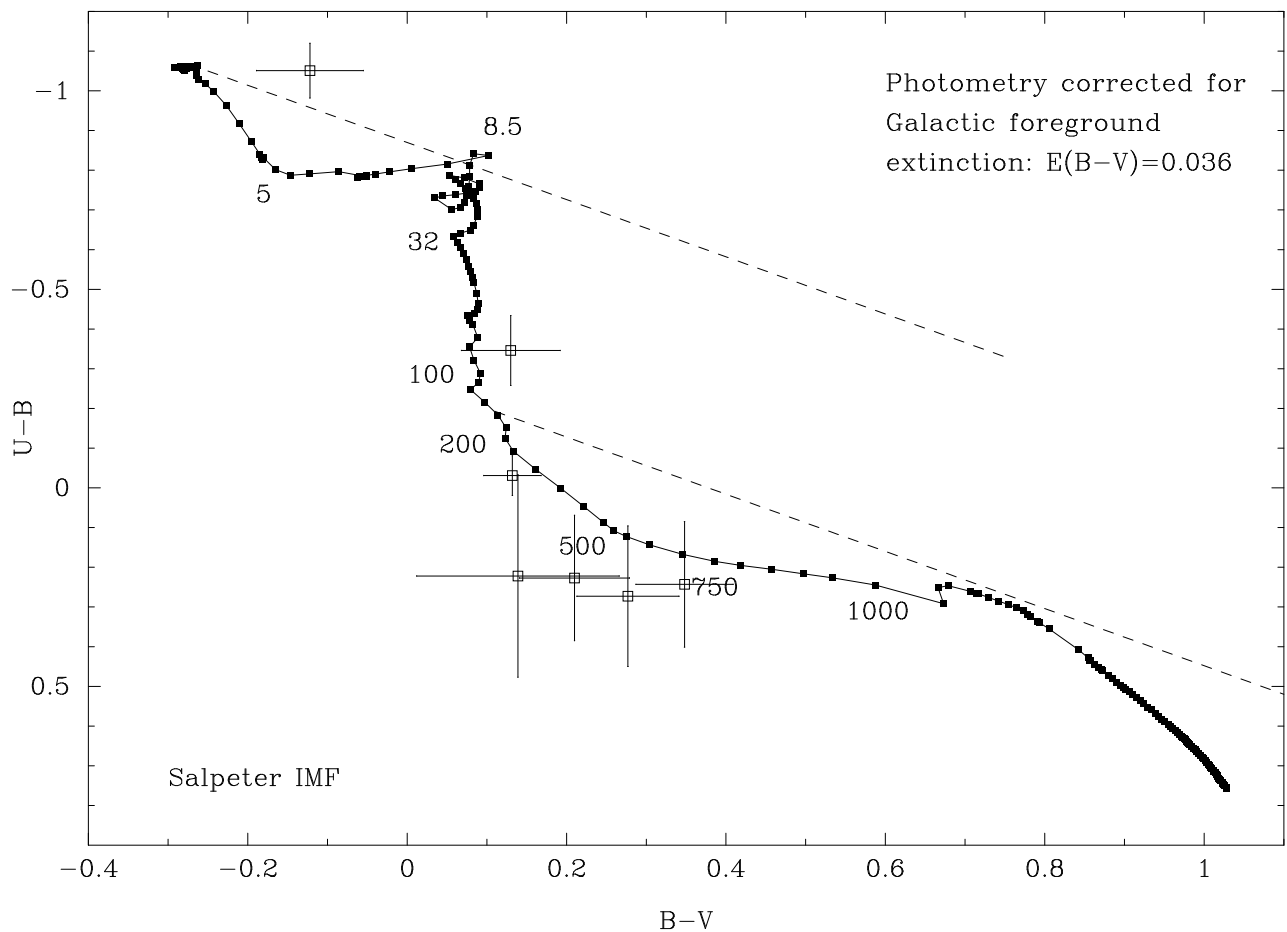}
\caption[f5.eps]{$U-B$ versus $B-V$ color-color diagram for the seven clusters
detected in $U$. 
Cluster models are from \citet{bc03} with a Salpeter initial mass function  
and solar-metallicity;  ages in Myr are indicated
for several models. The dashed lines
indicate reddening lines for a standard $R=3.1$ extinction law. The cluster
photometry has been corrected for a Galactic foreground reddening
$E(B-V) = 0.036$.
\label{ub-bv}}
\end{figure}

Figure~\ref{ub-bv} shows how the colors of our seven clusters with $U$
photometry compare with the \citet{bc03} models. 
Many of these clusters lie
in regions of the color-color plot where there is little
degeneracy between reddening and age. 
In particular, all the clusters
appear to be consistent with little or no additional reddening and 
ages of a few Myr up to several hundred Myr. 
The derived cluster masses,
ages, and reddenings are given in Table~\ref{tbl-mass1}.

The bluest cluster in this sub-sample is consistent with an age of 1-3 Myr
and a reddening $E(B-V) = 0.13-0.16$~mag from within Arp 220. 
Given the relatively central
location of this cluster, some additional reddening is not unreasonable.
Adopting an age of 3 Myr,
the derived cluster mass is $2.5\times 10^4 $ M$_\odot$. If the younger age
is more appropriate, the derived mass increases to $5 \times 10^4$ M$_\odot$.
In general, it is impossible to distinguish between these two ages for 
our youngest cluster candidates, as the model colors are so similar, and
so we give both mass and age estimates in Table~\ref{tbl-mass1}.

The remaining six clusters all agree with the model tracks without the
need to adopt any additional reddening and appear to have
intermediate ages of a few hundred Myr. Five of these clusters lie
in the outer regions of our image, which is again consistent with a lack
of additional reddening. (The sixth cluster lies very near the young, blue 
cluster discussed above, and so its apparent lack of reddening is somewhat
surprising.)
Five of these clusters have best-fit ages in the
range of 200-500 Myr and masses in the range of $3\times 10^5$ to 
$1.5\times 10^6$ M$_\odot$. The sixth cluster is one of the brightest in
our sample and has a best-fit age of 70 Myr and mass of $2\times 10^5$
M$_\odot$. 

The $\sim 100$ Myr ages of the six unreddened clusters strongly imply that
these are gravitationally bound star clusters and thus are intermediate-age
counterparts of the much older globular clusters. The long-term survival of
the youngest cluster cannot be predicted with any certainty.

\subsection{Clusters with NICMOS photometry\label{nicmos}}

The eight clusters identified in NICMOS images by \citet{s98} all
lie within our ACS/HRC field, although one of them 
(Scoville \#4) lies in the region shadowed by the occulting finger
(Figure~\ref{fig-color}).
For the remaining seven clusters, we can combine the NICMOS
photometry from \citet{s98,s00} with our $B-V$ and $V-I$ colors to place
strong constraints on the masses and ages of these clusters.

\begin{figure}
\includegraphics[angle=-90,scale=.7]{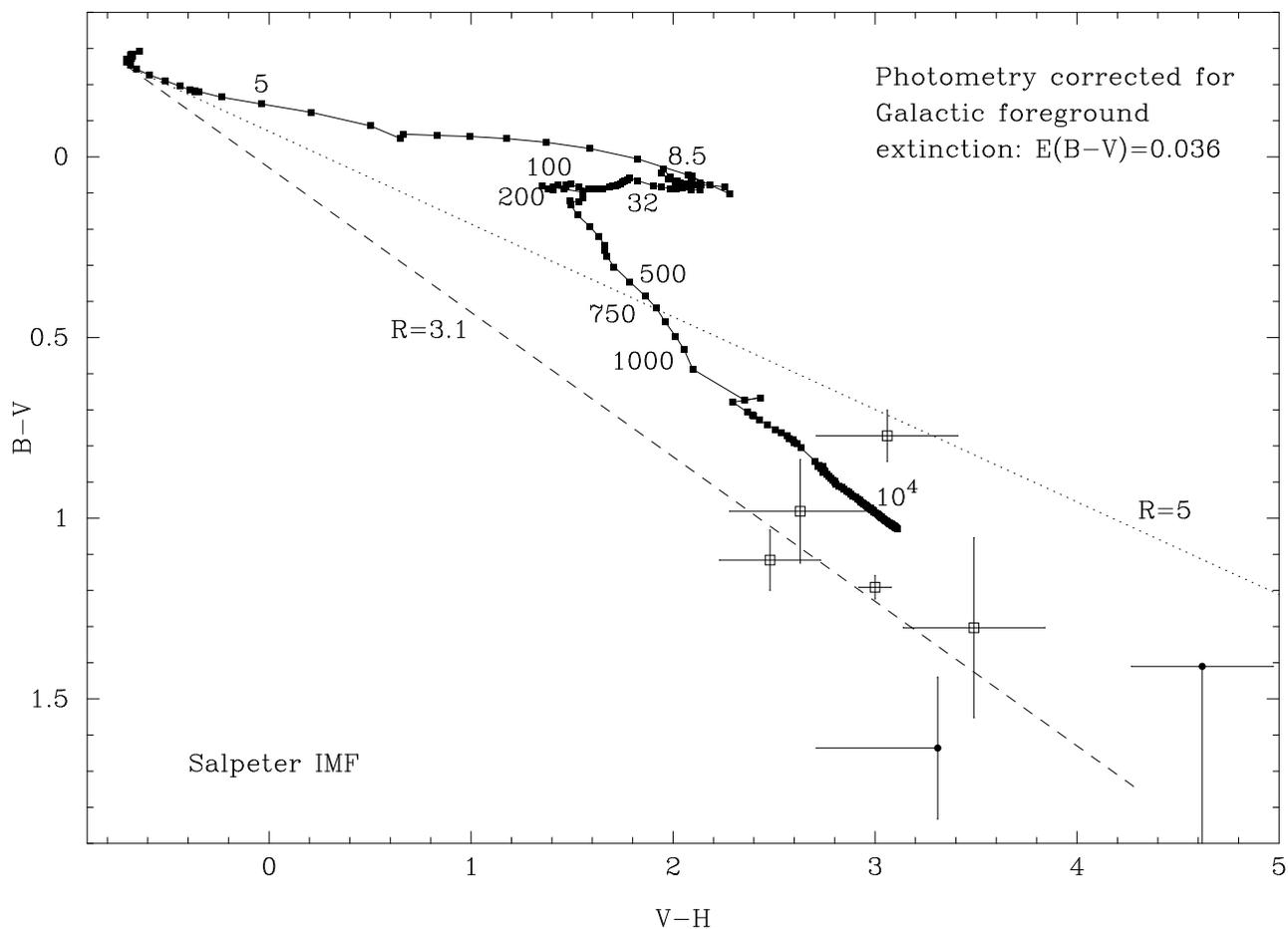}
\caption[f6.eps]{$B-V$ versus $V-H$ color-color diagram for the seven clusters
with published 1.6 $\mu$m photometry.. 
Cluster models are from \citet{bc03} with a Salpeter initial mass function  
and solar-metallicity; ages in Myr are indicated
for several models. The dashed line
indicates the reddening line for  1-3 Myr cluster with 
a standard $R=3.1$ extinction law; the dotted line indicates
the reddening line for $R=5$.\label{bv-vh}}
\end{figure}

Since only one of the clusters has an accurate $K$ magnitude \citep{s00},
we use the $B-V$ versus $V-H$ color magnitude diagram 
shown in Figure~\ref{bv-vh} to constrain
the cluster ages. The clusters are all extremely red
($V-H = 2.4-4.6$) and most lie below the theoretical cluster curve
in Figure~\ref{bv-vh}. The most natural interpretation is that these
are very young ($\sim 1-3$ Myr) clusters with significant
additional reddening.\footnote{Note that this conclusion is different from
that of \citet{s01}, who found ages of 10-100 Myr. However, their analysis
used first-generation WF/PC data that required deconvolution due to the error in
the figure of the HST primary mirror. We believe that our new data
give more reliable results.}
However, one cluster is  also consistent with an
intermediate age and reddening of 300 Myr and 0.6 mag, in which case
its mass would be $3\times 10^6$ M$_\odot$. This same cluster could
also be an unreddened globular cluster (see below).

We estimated masses for these clusters
by comparing their observed $V-H$ colors with the model $V-H$ colors
for clusters with ages of 1 Myr and 3 Myr.
The derived masses and reddenings 
are given in Table~\ref{tbl-mass1}. If the clusters have an age of 3  Myr,
then their masses range from $8\times 10^5$ to $6\times 10^6$ M$_\odot$,
substantially larger than the youngest cluster detected in the $U$ image
and, on average, larger than the masses of the intermediate-age clusters discussed
in the previous section. We note that one of the clusters may be
better fit by a non-standard extinction law with $R_v = 5$; although
this change would reduced the estimated $E(B-V)$ from 1.5 to 1 mag,
it would change the derived mass by only 15\%.

Two of the clusters have colors which are also consistent
with model colors appropriate to unreddened, 
13 Gyr old ``true'' globular clusters.
If these two clusters are extremely old, then
their masses would be $0.7-1\times 10^7$ M$_\odot$, comparable to the
most massive globular clusters found in giant elliptical galaxies
(\S 1). It is an interesting question whether we would expect a galaxy
such as Arp 220 to contain two such massive globular clusters; the fact
that M31 appears to possess a few globular clusters
in this range makes such a result at least possible.  

%**
Finally, we note that the brightest cluster in our sample (Scoville \#1)
is slightly non-stellar. Its observed profile, which is a convolution
of the intrinsic cluster profile and the PSF, has a FWHM
of 3.75 pixels in $I$, whereas the PSF alone has a mean FWHM of
3.0 pixels for the PSF (at a linear scale of 10 pc per pixel).
This comparison suggests that the true
half-light diameter of this object may very roughly be about 20 pc,
about five times larger than a normal globular cluster and
twice the size of even $\omega$ Cen. Interestingly, none of the
other objects in our list is noticeably nonstellar (i.e. broader than the
PSF), although for the fainter ones (and particularly those sitting on the
areas of complex background light) the distinction is harder to make.

\subsection{Clusters with only $B-V$ and $V-I$ colors}

\begin{figure}
\includegraphics[angle=-90,scale=.7]{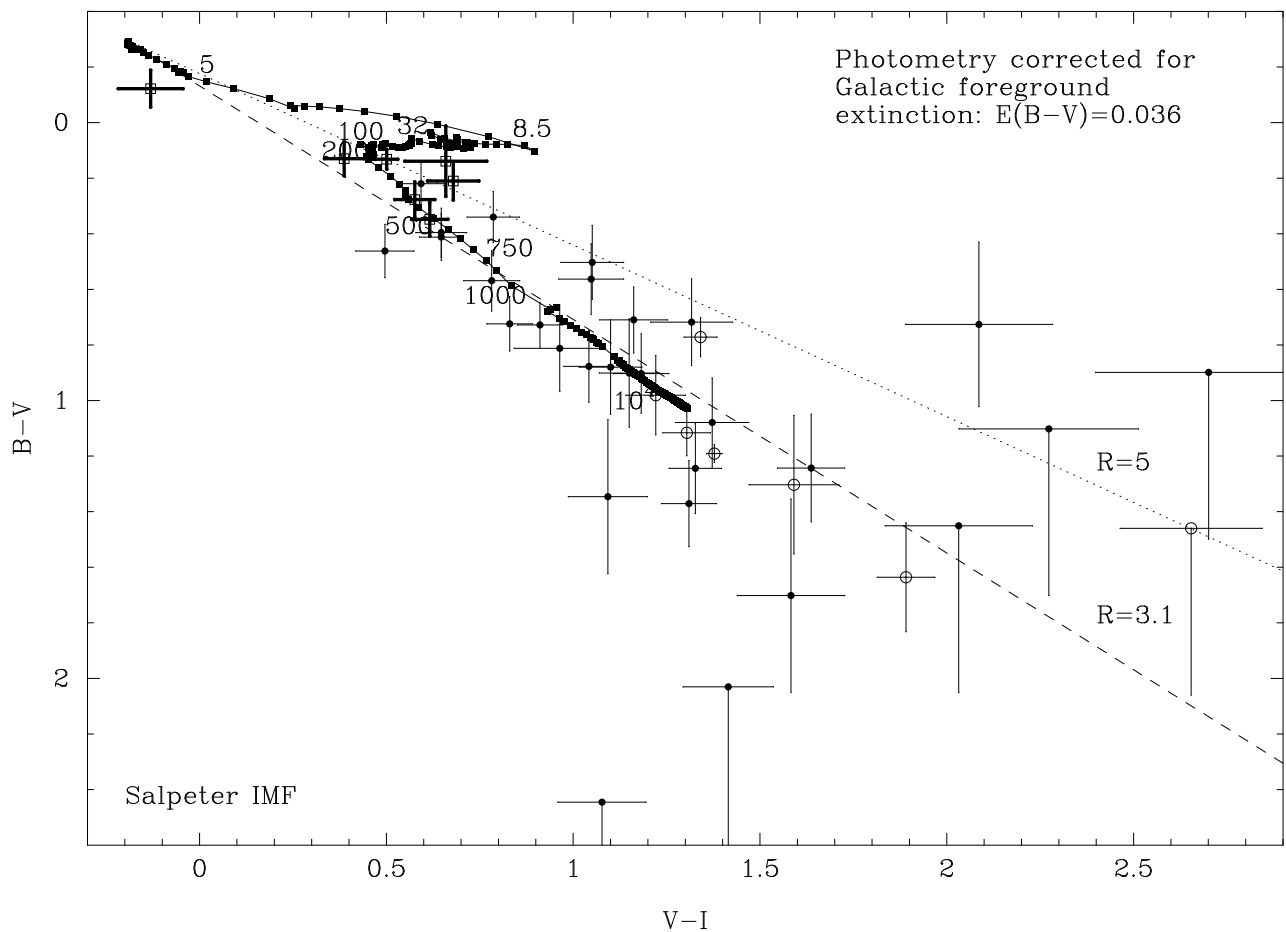}
\caption[f7.eps]{$B-V$ versus $V-I$ color-color diagram for clusters with $I< 24$ mag
(filled circles), clusters with $H$ photometry (open circles) and
clusters with $U$ photometry (open boxes).
Cluster models are from \citet{bc03} with a Salpeter initial mass function  
and solar-metallicity; ages in Myr are indicated
for several models. The dashed line
indicates the reddening line for  1-3 Myr cluster with 
a standard $R=3.1$ extinction law; the dotted line indicates
the reddening line for $R=5$.\label{bv-vi}}
\end{figure}

For objects with only $BVI$ data, there is a strong degeneracy between
age and reddening, particularly for ages greater than about 100 Myr.
Figure~\ref{bv-vi} shows the color-color diagram for all clusters in
our sample with $I < 24$ mag; the clusters detected in $U$ and $H$ are
also plotted. Upper limits to the $B-V$ color plotted for five of the clusters
are derived using the faintest detected $B$ magnitude in our sample; given
the variable background across our image, these upper limits should
be treated with caution.

There are a few clusters with very red $B-V$ and/or $V-I$ colors which lie in
the same region of the diagram occupied by the clusters detected
with NICMOS. These clusters seem likely to also be young, reddened clusters.
There are also a few clusters which lie above the model tracks with
$V-I \sim 1$ mag, which seem likely to be significantly reddened. However,
whether they are very young, reddened clusters, or intermediate-age clusters
with significant foreground reddening from other gas 
and dust in Arp 220, cannot
be determined at present.  Finally, there is one cluster with $V-I \sim 2$ mag 
and $B-V \sim
0.7$ mag that requires a relatively young age of $<10$ Myr and possibly
a non-standard reddening law to bring it into agreement with the
theoretical models. 

Even for the clusters with unusual
colors, there are always multiple combinations of age and
reddening which can fit the model colors.
However, it is possible to place some rough constraints on the masses
of the clusters simply by assuming that their ages lie between 1 Myr 
and 13 Gyr. The $BVI$ fluxes and colors of the evolutionary models
combine with the effects of reddening to 
vary in such a way that the mass of a given cluster can be constrained
to within a total range of a factor of $\sim 25$. Within this maximum
possible mass range, the largest masses correspond to an age of 13 Gyr,
and the smallest masses correspond to an age of 6 Myr, 
while  young (1 Myr) clusters lie near the middle of the mass
range. For the bluer clusters
in our sample for which the age is clearly $< 1$ Gyr, the mass can be
constrained more tightly, to within a factor of 8.

We have used this method to estimate mass ranges for each of the 24 clusters
with $BVI$ photometry from Table~\ref{tbl-phot}. The mass ranges are
given in Table~\ref{tbl-mass2} along with the mass that the cluster would
have if it had an age of 1 Myr. 
Clusters which have colors that could be consistent with 13 Gyr globular
clusters are noted in the comments to the table, as are clusters whose colors
imply they must be younger than 1 Gyr.
We use the mass calculated for an age of 1 Myr as the ``best'' mass
estimate, since it lies roughly in the middle of the mass range and makes
it easier to compare the properties of these clusters to the other young
clusters in Arp 220. (Masses smaller than this ``best'' mass only occur
for cluster ages from 3 Myr to $\sim 25$ Myr.)
These ``best'' masses range from a low of 
$1\times 10^5$ M$_\odot$ to a high of $4\times 10^6$ M$_\odot$.
The success of the NICMOS
images in picking out very young clusters suggests that deeper, 
high-resolution imaging of Arp 220 in the near-infrared could
help to identify additional young massive clusters and 
resolve some of the existing age ambiguities.

\subsection{The cluster spatial distribution}

We find that the objects in our study fall into two distinct age groups:
those less than $\sim 10$ Myr and those with intermediate ages
around 300 Myr.
The clusters that we have identified as ``young'' 
(with ages of 1-10 Myr) are clearly 
centrally concentrated. The 
center of the young cluster distribution is located about 1 kpc
east of the double nucleus of Arp 220 and the distribution
has an average radius of about 1.6 kpc. The intermediate-age clusters
are centered about 3 kpc to the north of the double nucleus
and their distribution has an average radius of about 3 kpc. However,
the intermediate-age clusters are found preferentially towards
the outskirts of our field (see Figure~\ref{xyplot}), while the 
field of view is not centered on the true nucleus of 
Arp 220. Thus, it is possible
that the true spatial distribution of the intermediate age clusters is
centered on the nucleus of Arp 220, but that we are missing intermediate-age
clusters at large radii, particularly to the south of the nucleus.

To investigate this issue further, we have calculated the radial
distribution of the various classes of clusters and cluster candidates
in our sample. We divided our field into three radial annuli 
($R<2.3$ kpc, $2.3 < R < 4.5$ kpc, and $4.5 < R < 6.8$ kpc) and corrected
the data for the incomplete areal coverage of the largest annulus. The
number of star clusters per square kiloparsec as a function of radius
is given in Table~\ref{tbl-radial}. All the cluster samples except the
intermediate-age clusters show a distinct radial gradient with the
largest numbers of clusters found closest to the nucleus of Arp 220.
This analysis suggests that even the very faint ($I > 24$ mag) objects
in our sample have a high probability of being star clusters in Arp 220,
as opposed to foreground stars or background objects.
The lack of central concentration of the intermediate-age clusters may
be understood by the fact that they all required $U$ detections for
good age estimates, which may prevent their identification
in the inner regions of the galaxy with large extinctions.

\section{Implications for Massive Star Cluster Formation\label{s4}}

The masses of the star clusters we have found
in Arp 220 are impressively large.
The intermediate-age clusters range from $2\times 10^5$ 
to $1.5\times 10^6$ M$_\odot$, within a factor of two of
the most massive clusters seen in the Antennae \citep{w99}.
The masses of some of the youngest star clusters may be even more
extreme. Depending on their exact age,
the most massive young clusters in Arp 220 may have masses as
large as $1\times 10^7$ M$_\odot$, comparable to the most massive
globular clusters seen in giant elliptical galaxies and the massive,
intermediate-age clusters seen in NGC 7252 and NGC 1316
\citep{maraston04,bastian06}.
Of course, their masses are expected to
decrease over the next 500 Myr due to the combined
effects of stellar mass loss, supernova-driven winds, and tidal
trimming in the central galaxy potential.

The age segregation seen in the spatial distribution of the clusters
suggests that the active region of cluster formation in Arp 220 was
larger 300 Myr ago than it is today.
Since the currently active region of star cluster formation in Arp 220
is roughly 1.6 kpc in radius, the intermediate-mass clusters
would have to have formed in a region roughly twice as large if
they formed {\it in situ}. 
NGC 7252 also shows a more compact spatial distribution for clusters 
younger than 10 Myr compared to the $\sim$ 300 Myr clusters which trace the 
overall light profile of the galaxy \citep{miller97}.
In contrast, the youngest star clusters in the Antennae 
are currently distributed over a region roughly 2.5 kpc
in size \citep{z01}, which is similar to the extent of the
intermediate-age clusters in Arp 220. Thus, it is plausible that
the earlier episode of massive cluster formation in Arp 220 had
a larger spatial extent than the current episode of cluster formation.

Our analysis shows that Arp 220 has experienced at least two recent
major episodes of massive star cluster formation, one around 300 Myr
ago and one in the last 5-10 Myr that is still continuing today.
The older clusters in our sample have ages
of 70-500 Myr, which is consistent with the estimated time
of $\sim 700$ Myr since the beginning of the 
interaction that produced Arp 220 
\citep{m01}. The average age of 300 Myr for these intermediate-age
clusters is in strikingly good agreement with the time at which
star formation is seen to increase in the prograde-retrograde model
 of \citet{mh96}
($t\sim 25$ in the dimensionless model units or $t \sim 400$ Myr if the 
progenitor galaxies have masses comparable to the Milky Way).
Other galaxies which show evidence for more than one episode of star
formation include the Antennae \citep{w99}, NGC 7252
\citep{miller97,maraston01}, and M51 \citep{bastian05}.

Given the fact that we cannot yet estimate accurate ages for most
of our cluster sample, it is possible that massive
star cluster formation in Arp 220 has actually been more continuous 
over the last 500 Myr than is apparent from these data.
It is striking that 
43-57\% of the clusters for which we have been able to determine
ages in our sample have ages of 10 Myr or less.
However, these numbers should be treated with caution since the
intermediate-age cluster sample, in particular, 
is likely very incomplete due to
the high and variable internal reddening in Arp 220.
It is also important to bear in mind that the mass ranges probed by
the young and intermediate-age cluster samples do not overlap
significantly. 
While the large number of very young clusters
seems to indicate an increase in cluster
formation activity in the last 10 Myr, it is unclear how many of these
clusters are gravitationally bound and likely to survive in the long
term. Indeed, the extremely high rate of cluster formation in the last
10 Myr that is seen in Arp 220 and in the Antennae \citep{zf99}
\citep[and to a lesser extent in M51,][]{bastian05}
strongly suggests that many of the observed young clusters in Arp 220 are
unbound and will dissipate well before reaching ages of 100 Myr or more.

Depending on their precise ages and masses,
the current star formation rate represented by the 7 most massive young 
clusters is 6-37 M$_\odot$ yr$^{-1}$. This is a significant fraction
(3-15\%) of the total current star formation rate in Arp 220 
[240 M$_\odot$ yr$^{-1}$, calculated from its far-infrared luminosity
\citep{s03} using the formula in \citet{k98}]. 
If the mass function of the star clusters is a power-law with slope -2,
then the total star formation rate in star clusters more massive than
$10^4$ M$_\odot$ would be 10-50\% of the current total star formation
rate in Arp 220. A similar calculation
for the intermediate-age clusters in our sample 
(assuming a cluster formation timescale of 100 Myr) 
gives a star formation rate 
of only 0.04 M$_\odot$ yr$^{-1}$ for the observed clusters
and 0.09 M$_\odot$ yr$^{-1}$ for clusters above $10^4$ M$_\odot$.
This calculation points to a substantially lower 
star formation rate in the earlier burst of star formation.
However, it is also possible that many of the young massive clusters do not
survive for more than a few tens of Myr, in which case
this calculation would understimate the true
star formation rate in clusters in the earlier burst. Finally, there
are large numbers of clusters in our sample for which we cannot determine an
accurate age, which could increase the estimated star formation
rates in one or both bursts. Assuming the same slope for the cluster
mass function, we would expect to find $\sim 70$ young clusters with
masses greater than $1-2 \times 10^5$ M$_\odot$ in Arp 220. 
This estimate suggests that
most of the star clusters in Table~\ref{tbl-mass2} are likely to be 
young clusters and that additional young massive clusters remain to
be identified in Arp 220, perhaps from the population with $I>24$ mag.

Returning to the comparison of Arp 220 with the Antennae, we can see
that the most massive cluster in Arp 220 is 2-3 times more massive than
the most massive cluster in the Antennae. In terms of the total number
of clusters, Arp 220 has 2-3 times as many clusters with masses above
$10^6$ M$_\odot$ as does the Antennae. Given the 25 times higher
star formation rate in Arp 220 compared to the Antennae, the number
of clusters in Arp 220 seems rather low. However, it is
important to keep in mind that the number of clusters identified in Arp 220
may be quite incomplete, even above $10^6$ M$_\odot$.
\citet{whit04} has suggested that the number of high-luminosity clusters
in starburst systems is predominantly a statistical effect of the
total cluster population present (see, for example, Figure 1 of his paper).
Unfortunately we cannot yet test this statement directly for
Arp 220, since the total number of clusters brighter than his suggested
fiducial level $M_V = -8$ cannot be established from our data.
If this size-of-sample effect is correct, then Arp 220 should have
very large numbers of young and moderately young clusters still to
be found, but these must be embedded in heavy and differential reddening.

\begin{figure}
\includegraphics[angle=-90,scale=.7]{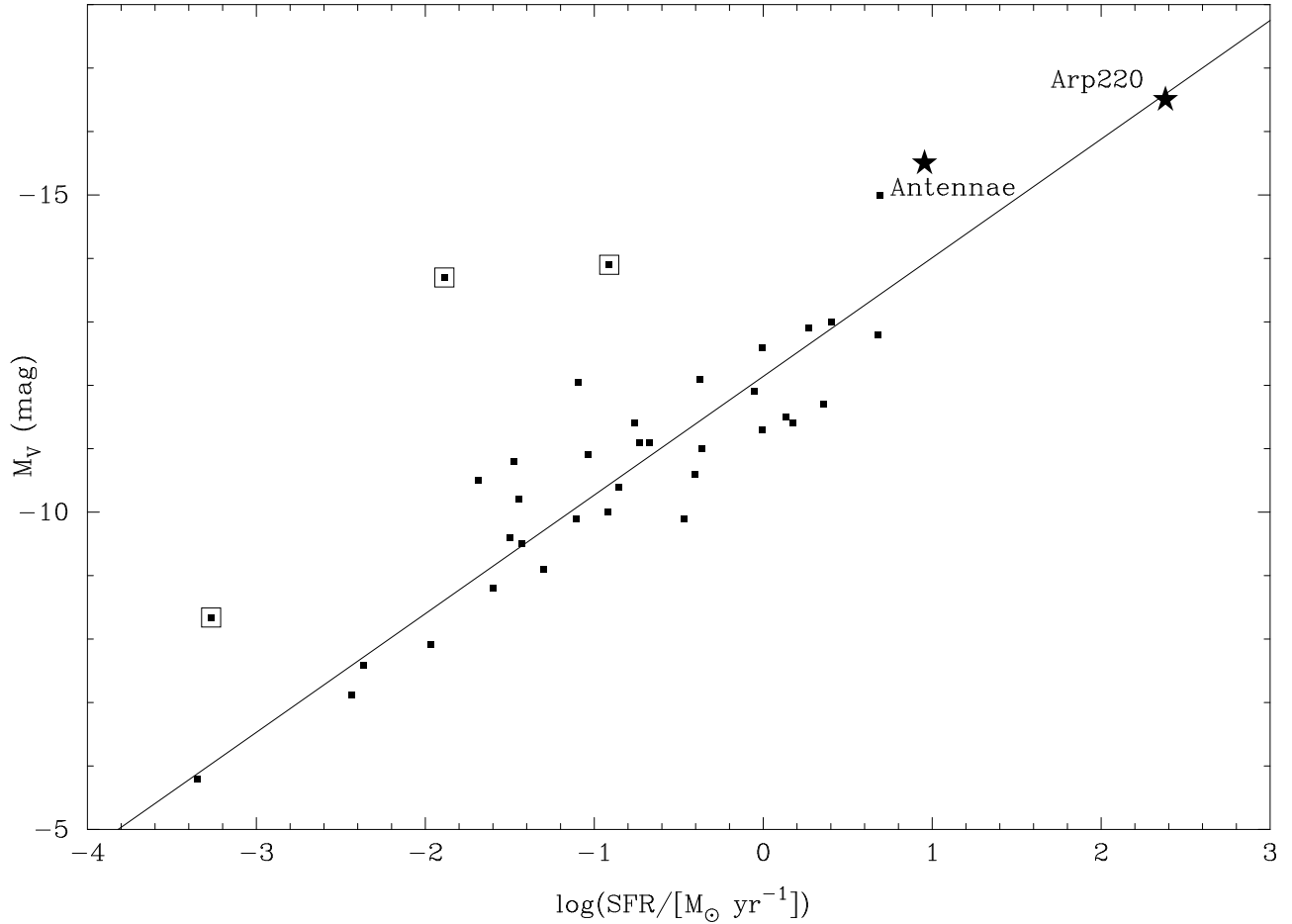}
\caption[f8.eps]{Brightest cluster magnitude, $M_v$, versus global
star formation rate for 36 galaxies from \citet{larsen02} with our results
for Arp 220 added. Data for Arp 220 and the Antennae are indicated by
the filled stars and data for three dwarf galaxies (DDO 165, NGC 1569,
and NGC 1705), which deviate significantly from the observed correlation,
are indicated by open squares. The line is the fit to 33 galaxies from
\citet{weidner04}. Including our new results for Arp 220 extends the
range of the observed correlation by more than one order of magnitude
in star formation rate.\label{sfr_mv}}
\end{figure}

An alternative approach is to compare the luminosity of the brightest cluster
with the total star formation rate \citep{bhe02,larsen02}. These two quantities
have been shown to be well-correlated in a wide variety of galaxies 
and this correlation 
has been suggested to be primarily a statistical effect. The correlation
appears to break down primarily for starburst dwarf galaxies
\citep{bhe02}, which are able to produce the occasional very massive
young cluster despite producing relatively few clusters overall.
Using the form of the correlation given in
\citet{weidner04},
it is clear that Arp 220 agrees very well with the relation derived
for galaxies with much lower star formation rates (Figure~\ref{sfr_mv}).
Interestingly,
the brightest cluster in the Antennae is about one magnitude too luminous
for its global star formation rate, which suggests that the cluster
formation process in the Antennae may be somewhat unusual.
The results for Arp 220 also suggest that the formation of the very massive
intermediate-age clusters seen in NGC 7252 and NGC1316 was probably 
accompanied by peak star formation rates in those galaxies in excess
of 100 M$_\odot$ yr$^{-1}$.

\section{Conclusions}

We have used new $UBVI$ optical imaging with the ACS/HRC camera on the
Hubble Space Telescope to identify 206 star cluster candidates in the
ultraluminous infrared galaxy Arp 220. 
These cluster candidates show 
a radial gradient in their surface density with distance from the
center of Arp 220, which suggests that most of them are star clusters
associated with the galaxy. One of the star clusters is spatially resolved
and may have a half-light diameter of roughly 20 pc, which would be 
twice the size of the massive Galactic globular cluster $\omega$ Cen.

Due to high and variable reddening,
only seven clusters are detected in our deep $U$ image.
We have been able to derive accurate
masses and ages for these seven clusters, as
well as for seven additional clusters with previously published 1.6 $\mu$m
data from the NICMOS camera. These clusters divide 
into two distinct age groups: young clusters with ages
$< 10$ Myr, and intermediate age clusters with ages of 70 to 500 Myr.
Most of the younger clusters are more massive than $10^6$ M$_\odot$, with
the most massive being perhaps as much as $10^7$ M$_\odot$ depending
on its precise age. The intermediate mass clusters are somewhat less massive
on average, ranging from $2\times 10^5$ to $2\times 10^6$ M$_\odot$.
Rough mass estimates for 24 clusters with $I < 24$ mag suggest most of
these clusters have masses in the range $10^5 - 10^6$ M$_\odot$.

The identification of a very young, massive star cluster in Arp 220 allows
us to extend the correlation between the global star formation rate and
the most luminous cluster seen by \citet{bhe02} by an order of magnitude.
This result implies that very high star formation rates are required
to form clusters more massive than $10^7$ M$_\odot$, which suggests that
the merger remnants NGC 7252 and NGC 1316 should have experienced peak star
formation rates greater than 100 M$_\odot$ yr$^{-1}$ at some point in
the merging process.

\acknowledgments

The research of C.D.W. and W.E.H. is supported by the Natural Sciences
and Engineering Research Council (Canada). C.D.W. also acknowledges support
from the Smithsonian Astrophysical Observatory.
The research of N.Z.S. is supported by NSF grant AST-0228955 and
by NASA under HST grant GO-09396.01-A.
The Space Telescope Science Institute
is operated by the Association of Universities for Research in Astrononmy,
Inc., for NASA under contract NAS5-26555.
We thank the referee for several useful comments that improved the
content of this paper.

Facilities: \facility{HST(ACS)}.

\clearpage

\clearpage

\clearpage

\clearpage

\clearpage

\clearpage

\clearpage

\begin{deluxetable}{cccc}
\tabletypesize{\scriptsize}
\tablecaption{Photometric Calibration Parameters\label{tbl-cal}}
\tablewidth{0pt}
\tablehead{
\colhead{Filter} & \colhead{Correction}  &\colhead{Zero}
&\colhead{Systematic} \\
\colhead{} 
& \colhead{to $r \rightarrow \infty$\tablenotemark{a}}
     &\colhead{Point\tablenotemark{b}}  
&\colhead{Uncertainty}
}
\startdata

F330W  & -0.42 & 22.904 & $\pm0.06$ \\

F435W  & -0.40 & 25.185 & $\pm0.04$ \\

F555W  & -0.44 & 25.255 & $\pm0.07$\\

F814W  & -0.65 & 24.849 & $\pm0.03$\\

\enddata

\tablenotetext{a}{Aperture correction in magnitudes from a radius of 3 pixels to
large radius, interpolated from  \citet{s05}.}

\tablenotetext{b}{For $UBVI$ on the VEGAMAG system, from \citet{s05}.}

\end{deluxetable}

\clearpage

\begin{deluxetable}{rrrcccccccc}
\tabletypesize{\scriptsize}
\tablecaption{ACS/HRC Photometry for Arp 220 Star Clusters \label{tbl-phot}}
\tablewidth{0pt}
\tablehead{ 
\colhead{ID} &  \colhead{x (px)} & \colhead{y (px)} & \colhead{RA (J2000) } &
\colhead{Dec (J2000) } & \colhead{$I$} & \colhead{$\sigma_I$} & \colhead{$(V-I)$} & 
\colhead{$\sigma_{V-I}$} & \colhead{$(B-V)$} & \colhead{$\sigma_{B-V}$} \\
}

\startdata
  1 &  814.6 &  632.0 & $15^h 34^m 57\fs14$ & $23^o 30'  6\farcs7$ &21.094 & 0.008 & 1.421 & 0.021 & 1.227 & 0.037 \\
  2 &  663.8 &  409.7 & 15 34 57.60 & 23 30  9.1 &21.893 & 0.051 & 1.347 & 0.064 & 1.152 & 0.083 \\
  3 &   55.3 &  836.7 & 15 34 57.08 & 23 30 26.4 &22.067 & 0.020 & 0.544 & 0.030 & 0.168 & 0.036 \\
  4 &  711.8 &  621.6 & 15 34 57.20 & 23 30  9.1 &22.416 & 0.022 & 1.934 & 0.077 & 1.672 & 0.196 \\
  5 &  777.0 &  526.8 & 15 34 57.34 & 23 30  7.1 &22.533 & 0.017 & 1.384 & 0.044 & 0.808 & 0.071 \\
  6 &  597.1 &  572.5 & 15 34 57.33 & 23 30 11.7 &22.761 & 0.045 &  ...  &  ...  &  ...  &  ...  \\
  7 &  785.4 &  580.9 & 15 34 57.24 & 23 30  7.2 &23.055 & 0.029 & 1.353 & 0.074 & 1.407 & 0.155 \\
  8 &  616.1 &  788.8 & 15 34 56.94 & 23 30 12.5 &23.130 & 0.032 & 0.954 & 0.060 & 0.764 & 0.080 \\
  9 &  617.7 &  706.4 & 15 34 57.09 & 23 30 12.0 &23.157 & 0.041 & 1.225 & 0.074 & 0.939 & 0.143 \\
 10 &  689.7 &  501.2 & 15 34 57.42 & 23 30  9.1 &23.163 & 0.046 & 1.415 & 0.098 & 1.115 & 0.160 \\
 11 &  435.5 &  682.7 & 15 34 57.20 & 23 30 16.3 &23.168 & 0.032 & 1.264 & 0.080 & 1.017 & 0.143 \\
 12 &  479.1 &  354.7 & 15 34 57.77 & 23 30 13.4 &23.263 & 0.025 & 2.697 & 0.190 &  ...  &  ...  \\
 13 &  671.0 &  364.4 & 15 34 57.67 & 23 30  8.8 &23.297 & 0.056 & 1.205 & 0.091 & 0.746 & 0.119 \\
 14 &  633.8 &  672.5 & 15 34 57.14 & 23 30 11.4 &23.314 & 0.060 & 1.360 & 0.110 & 0.754 & 0.156 \\
 15 &  595.1 &  739.3 & 15 34 57.04 & 23 30 12.7 &23.355 & 0.029 & 1.370 & 0.070 & 1.280 & 0.163 \\
 16 &  746.8 &  479.3 & 15 34 57.44 & 23 30  7.6 &23.361 & 0.031 & 0.690 & 0.058 & 0.448 & 0.084 \\
 17 &  897.5 &  520.8 & 15 34 57.31 & 23 30  4.1 &23.362 & 0.028 & 1.680 & 0.090 & 1.279 & 0.194 \\
 18 &  537.3 &  847.3 & 15 34 56.87 & 23 30 14.7 &23.381 & 0.027 & 0.659 & 0.049 & 0.384 & 0.061 \\
 19 &  569.2 &  780.1 & 15 34 56.98 & 23 30 13.6 &23.410 & 0.041 & 0.829 & 0.070 & 0.376 & 0.092 \\
 20 &  621.6 &  457.2 & 15 34 57.53 & 23 30 10.5 &23.415 & 0.027 & 1.634 & 0.120 & 1.339 & 0.249 \\
 21 &  815.4 &  527.7 & 15 34 57.32 & 23 30  6.2 &23.434 & 0.030 & 0.873 & 0.061 & 0.760 & 0.098 \\
 22 &  920.2 &  281.5 & 15 34 57.72 & 23 30  2.3 &23.443 & 0.027 & 0.690 & 0.068 & 0.432 & 0.088 \\
 23 &  678.8 &  315.3 & 15 34 57.76 & 23 30  8.2 &23.505 & 0.030 & 1.085 & 0.067 & 0.913 & 0.129 \\
 24 &   89.6 &  475.1 & 15 34 57.71 & 23 30 23.5 &23.513 & 0.035 & 0.619 & 0.054 & 0.313 & 0.064 \\
 25 &  622.3 &  879.3 & 15 34 56.78 & 23 30 12.8 &23.526 & 0.041 & 0.722 & 0.069 & 0.246 & 0.069 \\
 26 &  875.8 &  242.5 & 15 34 57.81 & 23 30  3.1 &23.536 & 0.028 & 1.193 & 0.080 & 0.937 & 0.195 \\
 27 &  710.9 &  533.4 & 15 34 57.36 & 23 30  8.7 &23.579 & 0.040 & 1.626 & 0.144 & 1.738 & 0.348 \\
 28 &  933.7 &  708.2 & 15 34 56.96 & 23 30  4.3 &23.589 & 0.039 & 1.094 & 0.084 & 0.539 & 0.133 \\
 29 &  899.6 &  592.1 & 15 34 57.18 & 23 30  4.4 &23.665 & 0.032 & 2.129 & 0.197 & 0.762 & 0.296 \\
 30 &  628.2 &  593.9 & 15 34 57.28 & 23 30 11.1 &23.723 & 0.041 & 2.744 & 0.303 &  ...  &  ...  \\
 31 &  131.4 &  958.3 & 15 34 56.84 & 23 30 25.2 &23.773 & 0.029 & 0.430 & 0.053 & 0.166 & 0.062 \\
 32 &  127.8 &  209.3 & 15 34 58.17 & 23 30 21.1 &23.777 & 0.042 & 1.143 & 0.084 & 0.916 & 0.170 \\
 33 &  637.1 &  484.6 & 15 34 57.48 & 23 30 10.3 &23.791 & 0.084 & 1.007 & 0.122 & 0.848 & 0.156 \\
 34 &  762.1 &  568.8 & 15 34 57.27 & 23 30  7.7 &23.801 & 0.050 & 1.120 & 0.118 &  ...  &  ...  \\
 35 &  739.7 &  324.2 & 15 34 57.72 & 23 30  6.9 &23.810 & 0.040 & 0.825 & 0.075 & 0.605 & 0.109 \\
 36 &  123.8 &  620.2 & 15 34 57.44 & 23 30 23.5 &23.839 & 0.044 & 0.539 & 0.078 & 0.498 & 0.096 \\
 37 &  580.7 &  371.1 & 15 34 57.70 & 23 30 11.0 &23.840 & 0.036 & 2.075 & 0.197 &  ...  &  ...  \\
 38 &  387.4 &  425.0 & 15 34 57.68 & 23 30 16.1 &23.878 & 0.043 & 1.458 & 0.121 &  ...  &  ...  \\
 39 &  246.5 &  757.8 & 15 34 57.15 & 23 30 21.3 &23.901 & 0.035 & 0.636 & 0.062 & 0.256 & 0.082 \\
 40 &  687.8 &  153.2 & 15 34 58.04 & 23 30  7.2 &23.908 & 0.031 & 1.136 & 0.106 & 1.382 & 0.277 \\
 41 &  685.2 &  779.5 & 15 34 56.93 & 23 30 10.7 &23.948 & 0.034 & 2.316 & 0.240 &  ...  &  ...  \\
 42 &  542.8 &  258.6 & 15 34 57.91 & 23 30 11.3 &23.960 & 0.045 & 1.091 & 0.087 & 0.599 & 0.127 \\
 62 &  508.7 &  408.5 & 15 34 57.66 & 23 30 13.0 &24.274 & 0.067 & 0.702 & 0.110 & 0.175 & 0.127 \\
 86 &  514.8 &  401.8 & 15 34 57.54 & 23 30 12.8 &24.590 & 0.073 &-0.088 & 0.087 &-0.086 & 0.067 \\

\enddata

\tablecomments{The complete version of this table is in the electronic edition 
of the Journal.  The printed edition contains only a sample.}

\end{deluxetable}

%\clearpage

\begin{deluxetable}{cccccccccccccl}
\tabletypesize{\scriptsize}
\tablecaption{Masses and Ages for Clusters in Arp 220\label{tbl-mass1}}
\tablewidth{0pt}
\tablehead{
\colhead{ID} 
& \colhead{$V$} & \colhead{$\sigma_V$}   &\colhead{$B-V$} &\colhead{$\sigma_{B-V}$}
   & \colhead{$U-B$} & \colhead{$\sigma_{U-B}$}    &
\colhead{$V-H$\tablenotemark{a}} & \colhead{$\sigma_{V-H}$\tablenotemark{a}} & \colhead{age} & \colhead{$E(B-V)$} & \colhead{mass} & \colhead{Scoville}  \\
\colhead{} 
& \colhead{} & \colhead{} & \colhead{} & \colhead{}  & 
\colhead{}     &\colhead{}    &
\colhead{} & \colhead{} & \colhead{(Myr)} & \colhead{} & \colhead{(M$_\odot$)} & \colhead{ID\tablenotemark{a}}
}
\startdata
1 & 22.515 & 0.019 & 1.227 & 0.037 & ... & ... & 3.09 & 0.08 & 1-3\tablenotemark{b} & 1.48 & $0.6-1.2\times 10^7$ & 1  \\
2 & 23.240 & 0.039 & 1.152 & 0.083 & ... & ... & 2.57 & 0.25 & 1-3\tablenotemark{b} & 1.27 & $2-4\times 10^6$ & 2  \\
3  & 22.611 & 0.022 & 0.168 & 0.036 & -0.005 & 0.050 & ... & ... & 200 & 0 & $1.5\times 10^6$ & ...\\
4 & 24.350 & 0.074 & 1.672 & 0.196 & ... & ... & $<$3.40 & ... & 1-3\tablenotemark{bc} & 1.71 & $2-4 \times 10^6$ &  3  \\
5 & 23.917 & 0.041 & 0.808 & 0.071 & ... & ... & 3.15 & ... & 1-3\tablenotemark{bd} & 0.96 & $2-4\times 10^6$& 5  \\  
11 & 24.432 & 0.073 & 1.017 & 0.143 & ... & ... & 2.72 & ... & 1-3\tablenotemark{be} & 1.33 & $0.8-1.6\times 10^6$ & 7 \\  
12 & 25.960 & 0.188  & ... & ... & ... & ... & 4.71 & ... & 1-3\tablenotemark{b} & 2.13 & $2-4\times 10^6$ & 8 \\  
18 & 24.040 & 0.041 & 0.384 & 0.061 & 0.269 & 0.158 & ... & ... & 500 & 0 & $7\times 10^5$ & ... \\
20 & 25.049 & 0.117 & 1.339 & 0.249 & ... & ... & 3.58 & ... & 1-3\tablenotemark{b} & 1.68 & $1-2\times 10^6$ & 6  \\  
24 & 24.132 & 0.041 & 0.313 & 0.064 & 0.299 & 0.177 & ... & ... & 400 & 0 & $5\times 10^5$ & ... \\
25 & 24.248 & 0.055 & 0.246 & 0.069 & 0.253 & 0.158 & ... & ... & 500 & 0 & $5\times 10^5$ & ... \\
31 & 24.203 & 0.044 & 0.166 & 0.062 & -0.320 & 0.088 & ... & ... & 70 & 0 & $2\times 10^5$ & ... \\
62 & 24.976 & 0.087 & 0.175 & 0.127 & 0.248 & 0.255 & ... & ... & 400 & 0 & $3\times 10^5$ & ... \\
86 & 24.502 & 0.047 & -0.086 & 0.067 & -1.025 & 0.069 & ... & ... & 1-3\tablenotemark{b} & 0.15 & $2.5-5\times 10^4$ & ... \\
 \enddata

\tablecomments{A distance to Arp 220 of 77 Mpc is assumed throughout.
Masses are derived from \citet{bc03} models assuming a Salpeter initial
mass function and a standard reddening law (see text).}

\tablenotetext{a}{NICMOS 1.6 $\mu$m photometry from \citet{s98} (Clusters 1, 2, 4) and
\citet{s00} (Clusters 5, 11, 12, 20). Cluster identification number from \citet{s98}.}

\tablenotetext{b}{It is impossible to distinguish between these two young
ages; the older age of 3 Myr corresponds to the smaller mass.}

\tablenotetext{c}{Since this cluster has only an upper limit to $V-H$, its 
reddening was estimated from the $V-I$ color.}

\tablenotetext{d}{Another possible solution is an unreddened 13 Gyr
cluster with mass of $1\times 10^7$ M$_\odot$. A third possible  solution
is a 300 Myr cluster with $E(B-V)=0.56$ and mass $3\times 10^6$ M$_\odot$.}

\tablenotetext{e}{Another possible solution is an unreddened 13 Gyr
cluster with mass of $7\times 10^6$ M$_\odot$.}

\end{deluxetable}

\begin{deluxetable}{rccl}
\tabletypesize{\scriptsize}
\tablecaption{Mass Estimates for Additional Clusters in 
Arp 220\label{tbl-mass2}}
\tablewidth{0pt}
\tablehead{
\colhead{ID} 
& \colhead{Mass\tablenotemark{a}} & \colhead{Mass Range}   
&\colhead{Comments} \\
\colhead{} 
& \colhead{(M$_\odot$)} & \colhead{(M$_\odot$)} & \colhead{} 
}
\startdata
7 & $4.0\times 10^6$ & $(1-20)\times 10^6$ & \\
8 & $9\times 10^5$ & $(2-50)\times 10^5$ & \\
9 & $1.0\times 10^6$ & $(3-70)\times 10^5$ & possible old globular cluster\\
10 & $1.6\times 10^6$ & $(4-90)\times 10^5$ & \\
13 & $5\times 10^5$ & $(1-30)\times 10^5$ & \\
14 & $5\times 10^5$ & $(1-30)\times 10^5$ & \\
15 & $2.1\times 10^6$ & $(0.5-10)\times 10^6$ & \\
16 & $4\times 10^5$ & $(0.9-7)\times 10^5$ & age $<1$ Gyr\\
17 & $1.6\times 10^6$ & $(0.4-10)\times 10^6$ & \\
19 & $2\times 10^5$ & $(0.6-5)\times 10^5$ & age $<1$ Gyr\\
21 & $7\times 10^5$ & $(2-40)\times 10^5$ & \\
22 & $3\times 10^5$ & $(0.8-7)\times 10^5$ & age $<1$ Gyr\\
23 & $8\times 10^5$ & $(2-50)\times 10^5$ & possible old globular cluster\\
26 & $8\times 10^5$ & $(2-50)\times 10^5$ & possible old globular cluster\\
27 & $4.9\times 10^6$ & $(1-30)\times 10^6$ & \\
28 & $3\times 10^5$ & $(0.7-20)\times 10^5$ & \\
29 & $2\times 10^5$ & $(0.4-10)\times 10^5$ & \\
32 & $6\times 10^5$ & $(2-40)\times 10^5$ & possible old globular cluster\\
33 & $6\times 10^5$ & $(1-30)\times 10^5$ & possible old globular cluster\\
35 & $3\times 10^5$ & $(0.8-7)\times 10^5$ & age $<1$ Gyr\\
36 & $3\times 10^5$ & $(0.8-6)\times 10^5$ & age $<1$ Gyr\\
39 & $1\times 10^5$ & $(0.3-3)\times 10^5$ & age $<1$ Gyr\\
40 & $2.1\times 10^6$ & $(0.5-10)\times 10^6$ & possible old globular cluster\\
42 & $2\times 10^5$ & $(0.6-10)\times 10^5$ & \\
 \enddata

\tablecomments{A distance to Arp 220 of 77 Mpc is assumed throughout.
Masses are derived from \citet{bc03} models assuming a Salpeter initial
mass function and a standard reddening law (see text).}

\tablenotetext{a}{Mass calculated assuming age of 1 Myr (see text).}

\end{deluxetable}

\begin{deluxetable}{cccccc}
\tablecaption{Radial Distribution of Clusters in Arp 220\label{tbl-radial}}
\tablewidth{0pt}
\tablehead{
\colhead{Annulus} 
& \colhead{Age $<$} & \colhead{Age 70--}   &\colhead{Clusters with } &\colhead{Clusters with}
   & \colhead{All cluster} \\
\colhead{(kpc)} 
& \colhead{10 Myr} & \colhead{500 Myr} & \colhead{$I<24$ mag} & \colhead{$I>24$ mag}  & 
\colhead{candidates}     
}
\startdata
$R<2.3$  &       0.38    & 0  &   0.94  &   2.5   &     3.8 \\
$2.3<R<4.5$ &   0.17 &  0.06  &   0.21  &  1.7   &     2.1 \\
$4.5 < R < 6.8$ & 0  &   0.06 &  0.11 &  0.7     &   0.9 \\
 \enddata

\tablecomments{Units are number of clusters per square kiloparsec.}

\end{deluxetable}

\end{document}